\documentclass[doublecol]{epl2} 

\usepackage{graphicx} 
\usepackage{dcolumn}
\usepackage{bm}
\usepackage{amssymb,amsmath}
\usepackage{epstopdf}
\usepackage{color}

\newcommand{\beq}{\begin{equation}}
\newcommand{\beqn}{\begin{eqnarray}}
\newcommand{\eeq}{\end{equation}}
\newcommand{\eeqn}{\end{eqnarray}}
\newcommand{\nn}{\nonumber}

\newcommand{\me}{m_{\rm e}}
\newcommand{\mh}{m_{\rm h}}
\newcommand{\ome}{\omega_{\rm e}}
\newcommand{\omh}{\omega_{\rm h}}
\newcommand{\omp}{\omega^{\prime}}

\newcommand{\bv}{{\bf v}}

\newcommand{\bq}{{\bf q}}

\newcommand{\be}{{\bf e}}

\newcommand{\Bmath}{\mathcal{B}}
\newcommand{\Amath}{\mathcal{A}}

\title{Measure of Diracness in two-dimensional semiconductors}

\author{M. O. Goerbig\inst{1}, G. Montambaux\inst{1}, F. Pi\'echon\inst{1}}
\shortauthor{M. O. Goerbig, G. Montambaux, F. Pi\'echon}

\institute{
\inst{1} Laboratoire de Physique des Solides, %
Univ.~Paris-Sud, CNRS UMR 8502, F-91405 Orsay, France
}

\date{\today}

\abstract{
We analyze the low-energy properties of two-dimensional direct-gap semiconductors, such as for example the transition-metal dichalcogenides MoS$_2$, WS$_2$, and their
diselenide analogues MoSe$_2$, WSe$_2$, etc., which are currently intensively investigated. In general, their electrons have a mixed character -- they can
be massive Dirac fermions as well as simple Schr\"odinger particles. We propose a measure (Diracness) for the degree of mixing between the two characters and discuss
how this quantity can in principle be extracted experimentally, within magneto-transport measurements, and numerically via \textit{ab initio} calculations.
}


\begin{document}

\maketitle

\section{Introduction}

Graphene research has triggered an enormous amount of work in the understanding of Dirac fermions in two-dimensional (2D) materials.
A second generation of 2D crystals is now available, e.g., in the form of exfoliated boron nitride and transition-metal dichalcogenides
\cite{NovoselPNAS}, such
as molybdenum disulfide (MoS$_2$) \cite{makPrl,exp1} or tungsten disulfide (WS$_2$) \cite{ZengSR,WangNT}, as well as their diselenide analogues
\cite{TongayNL,Tonndorf,zhaoNano}.
Whereas the low-energy electrons in graphene may be described in terms of massless Dirac fermions, the
situation is different in most of the abovementioned second-generation 2D crystals, which are direct-gap
semiconductors \cite{mattheiss,lebPrb,yunPrb,cheiPrb,rosArx,roldanArx,ochArx}.
Although
the latter have been modeled in several studies as massive Dirac fermions \cite{roldanArx,xiaoPrl,rose}
a clear identification of the Dirac character, or \textit{Diracness},
is yet lacking.
A complication in this identification is certainly the rather large band gap, which limits the energy window where
one may hope to describe the low-energy electrons in terms of massive Dirac fermions. Furthermore, if only the parabolicity of the conduction
and the valence band is taken into account, in addition to the band gap, there is a large amount of arbitrariness in modeling the band structure.
Instead of appealing to Dirac fermions, the band structure can conveniently be modeled in terms of two types of
Schr\"odinger fermions with the appropriate band mass that yield two completely decoupled bands. One needs to point out that the coupling between
the bands is not a consequence of the band gap or its absence -- for the same gap value, the bands are strongly coupled when modeled as massive Dirac
fermions, whereas they are decoupled for Schr\"odinger fermions. Indeed, the Diracness is encoded in the
Hamiltonian, and it cannot be extracted from knowledge about the band energies alone.

The Diracness, which we propose to quantify in this paper, is nevertheless a physical observable that is unveiled by a magnetic field $B$. If
one considers the two parabolic bands, Landau quantization yields a spectrum $\epsilon_{{\rm e},n}\simeq \Delta + \hbar (eB/\me^C)(n+\gamma_{\rm e})$
for electrons in the conduction band and $\epsilon_{{\rm h},n}\simeq -\Delta - \hbar (eB/\mh^C)(n+\gamma_{\rm h})$ in the valence band, in terms of the integer $n$.
Here, the parameter $\Delta$ yields the direct gap $2\Delta$, and $\me^C$ and $\mh^C$ are the electron (valence band) and the hole
(conduction band) cyclotron masses, respectively. The phases $\gamma_{\rm e}$ and $\gamma_{\rm h}$ have mainly been described within semiclassical approaches
\cite{roth,wilkenson,mikitik}.
They are related to the Berry phase and may be extracted from Shubnikov-de-Haas oscillations measured in magneto-transport experiments.
Whereas Schr\"odinger fermions yield phases $\gamma_{\rm e}=\gamma_{\rm h}=1/2$, massive Dirac fermions are strikingly different and
the calculation of the Landau-level spectrum yields a phase $\gamma_{\rm e}=\gamma_{\rm h}=0$ \cite{haldane,fuchs}. A recent theoretical study on surface states
of three-dimensional topological insulators in a magnetic field
has shown that this phase is not necessarily quantized due to electron-hole asymmetry
and can vary between the two limits \cite{wright}.

Here, we propose to quantify the notion of Diracness, which may also be viewed as the degree of inter-band coupling,
for electrons in 2D direct-gap semiconductors. In a first step, we consider a simplified model that unveals
the basic features of the Diracness in a transparent manner. The parameter is identified in an expansion of the Landau-level spectrum as well
as in a semiclassical treatment that allows for a direct connection between the phase offset and the Berry phase.
Afterwards we discuss the Diracness in the most general model, within the Luttinger-Kohn representation \cite{LKrep} that accounts naturally for
the different possible band anisotropies, in a second part. The measure we introduce
may be viewed as the contribution of the Dirac mass $m_D=\Delta/v_D^2$, in terms of the direct
gap $\Delta$ and the Dirac velocity $v_D$, to the overall band mass, which is extracted from the band structure.
In contrast to this band mass, special efforts need to be invested in obtaining the Dirac mass although it is a
physical observable. Indeed, it can be extracted experimentally from Shubnikov-de-Haas measurements and
numerically, via \textit{ab initio} calculations, by a determination of the Berry curvature at the direct gap.

\section{Diracness in a simplified model}

Let us first consider the basic structure of a direct-gap semiconductor, which is that of a conduction band with a curvature
given by the mass $m_{\rm e}$ and of a valence band with mass $m_{\rm h}$. For the sake of simplicity, we consider the masses to
be isotropic here, whereas an anisotropy in the parameters naturally arises in the general Luttinger-Kohn representation discussed
later on. The bands are separated by a gap $2\Delta$ at an arbitrary point $K_0$ in reciprocal space. 
Notice that, in the case of a time-reversal-symmetric system,
there is another point $K_0'=-K_0$ where the band structure also reveals a direct gap unless $K_0$ is
itself a time-reversal-invariant momentum
in the first Brillouin zone of the underlying lattice, in which case the the Hamiltonian has a completely different structure
\cite{montambaux}. Eventually, due to spin-orbit coupling the gap
$\Delta$ can be spin-dependent, but this does not alter the validity of the following argumentation such that we omit
this case for simplicity.

The situation can be captured within a simple two-band model described by the Hamiltonian 
\beq\label{eq:ham0}
H=\left(\begin{array}{cc}
\Delta + \frac{\hbar^2q^2}{2\me^0} & \hbar v_D(q_x -iq_y)\\
\hbar v_D(q_x+iq_y) & -\Delta -\frac{\hbar^2 q^2}{2\mh^0}
\end{array}
\right)
\eeq
in the vicinity of one of the points $K_0$.
One retrieves the limit of Schr\"odinger fermions and decoupled bands for vanishing off-diagonal terms (Dirac velocity $v_D=0$) \cite{note1},
whereas the model represents massive Dirac fermions for $1/\me^0=1/\mh^0=0$.
Notice furthermore that the masses $\me^0$ and $\mh^0$ coincide with the band masses $\me$ and $\mh$ only for a vanishing Dirac
velocity. Indeed, the electronic spectrum of Hamiltonian (\ref{eq:ham0}) reads
\beq\label{eq:disp00}
\epsilon_{\lambda}(q)=\frac{\hbar^2q^2}{2\mu} +\lambda \sqrt{\hbar^2v_D^2q^2 +\left(\Delta + \frac{\hbar^2q^2}{2M}\right)^2},
\eeq
where $\lambda=+$ for the conduction band of electrons (e) and $\lambda=-$ for the valence band of holes (h). The parameters
$\mu$ and $M$ are related to the bare masses via $1/\me^0=1/M + 1/\mu$ and $1/\mh^0 = 1/M - 1/\mu$.
A simplified model, for $1/M=0$ i.e. $\me^0=-\mh^0$, has recently been investigated in Refs. \cite{wright,wang} to model surface states
of three-dimensional topological insulators.

Due to the off-diagonal
terms in Hamiltonian (\ref{eq:ham0}), the bands are not simply parabolic. However, corrections beyond parabolicity may also arise from
additional terms that we have neglected in our model. For the sake of consistency, we need to restrict ourselves to the spectrum
expanded to second order in $q$,
\beq\label{eq:disp0}
\epsilon_{\lambda}(q)\simeq \lambda\left(\Delta +\frac{\hbar^2 q^2}{2m_{\lambda}}\right),
\eeq
where the band masses read
\beq\label{eq:bandM}
\frac{1}{m_{\lambda}}=\frac{1}{m_{\lambda}^0}+\frac{1}{m_D},
\eeq
in terms of the Dirac mass $m_D=\Delta/v_D^2$. One clearly notices that the band structure does not allow one to identify the four
relevant parameters of the model ($\me^0$, $\mh^0$, $v_D$ and $\Delta$) but only $\me$, $\mh$, and $\Delta$ -- one could thus always
retrieve the band structure with a simple model of decoupled bands for $v_D=0$, without appealing to a Dirac character (Diracness)
of the charge carriers.

However, the Diracness is revealed in two physical quantities: the Berry curvature and the phase offset of the Shubnikov-de-Haas oscillations,
as we show below. A semiclassical treatment shows that both quantities are intertwined.

\subsection{Landau-level spectrum}

Before calculating the Berry curvature and the phase offset in a semiclassical approach, we show in this section how the Diracness can be extracted from the
Landau-level spectrum associated with our Hamiltonian (\ref{eq:ham0}), via the Peierls substitution $(q_x-iq_y)\rightarrow \sqrt{2}a/l_B$. Here,
$l_B=\sqrt{\hbar/eB}\simeq 26$ nm$/\sqrt{B{\rm [T]}}$ is the magnetic length, and $a$ is the usual harmonic-oscillator ladder operator, which
satisfies $[a,a^{\dagger}]=1$. The Hamiltonian 
thus reads
\beq\label{eq:hamB}
H_B=\left(\begin{array}{cc}
\Delta +\hbar\ome^0(a^{\dagger}a + 1/2) & \hbar\omp a\\ \hbar\omp a^{\dagger} & -[\Delta + \hbar\omh^0(a^{\dagger}a + 1/2)]
\end{array}\right),
\eeq
where the cyclotron-type frequencies are $\omega_{\lambda}^0=eB/m_{\lambda}^0$ 
and $\omp=\sqrt{2}v_D/l_B$. Hamiltonian (\ref{eq:hamB}) can be diagonalized with the help of the ansatz
\beqn\nn
\psi_n &=& \left(\begin{array}{c} \alpha |n-1\rangle \\ \beta|n\rangle \end{array}\right) \qquad {\rm for} ~n>0 \\
{\rm and} \qquad
\psi_0 &=& \left(\begin{array}{c} 0 \\ |0\rangle \end{array}\right) \qquad {\rm for} ~n=0.
\eeqn
The zero mode ($n=0$) has a negative energy,
\beq\label{eq:LL0}
\epsilon_{n=0}=-\left(\Delta + \frac{\hbar\omh^0}{2}\right),
\eeq
and is thus situated in the valence band. This is a manifestation of the so-called parity anomaly, common for relativistic field theories in even space
dimensions \cite{Semenoff}, and the presence of a second point ($K_0^{\prime}$) in the first Brillouin zone ensures a second zero mode in the conduction band
at $\epsilon_{n=0}^{\prime}=\Delta + \hbar\ome^0/2$.

The $n\neq 0$ Landau levels are also readily obtained by solving the eigenvalue equation
$H_B\psi_{n}=\epsilon_n \psi_{n}$ for $\alpha$ and $\beta$, and one obtains two sets of levels (for the bands $\lambda=\pm$) with
\beq\label{eq:LLn}
\frac{\epsilon_{\lambda,n}}{\hbar}=\delta\omega n -\frac{\Omega}{2} +\lambda \sqrt{\left(\frac{\Delta}{\hbar} + \Omega n -\frac{\delta\omega}{2}\right)^2 + {\omp}^2 n},
\eeq
in terms of the frequencies $\Omega=eB/M$ and $\delta\omega = eB/\mu$, which are related to the abovementioned cyclotron-type frequencies by
$\ome^0=\Omega+\delta\omega$ and $\omh^0=\Omega -\delta\omega$. Notice that this result also includes the expression (\ref{eq:LL0}) for the energy of
the zero mode for $n=0$ and $\lambda=-$. An expression similar to (\ref{eq:LLn}) has recently been obtained for the Landau-level spectrum in two-dimensional
molybdenum disulfide, where the terms $\delta\omega$ and $\Omega$ were treated perturbatively \cite{rose}. Furthermore, Eq. (\ref{eq:LLn}) coincides,
for $\Omega=0$ and the gap given by the Zeeman effect, with the Landau-level spectrum calculated for surface states of Bi$_2$Se$_3$ \cite{wang}.

\begin{figure}
\centering
\includegraphics[width=8.5cm,angle=0]{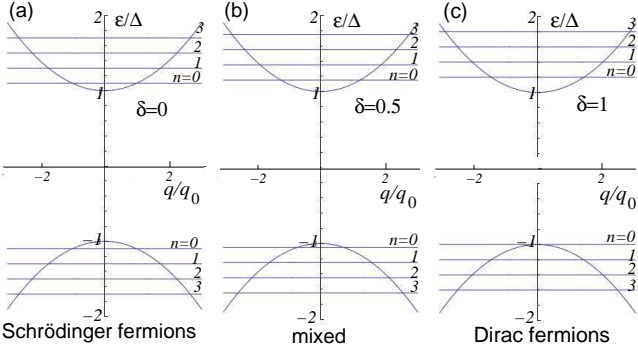}
\caption{\footnotesize{Landau levels for different values of the Diracness
[(a) $\delta=0$, (b) $\delta=0.5$, and (c) $\delta=1$], for an electron-hole-symmetric situation ($1/\mu=0$).
The energy is measured in units of the gap parameter $\Delta$ and the wave vectors in units of $q_0=\sqrt{2M\Delta}/\hbar$
(in the sketch of the underlying $B=0$ curves). We have chosen 
a value of $\hbar\omega/\Delta=0.2$ for the cyclotron frequency.
}}
\label{fig1}
\end{figure}

As in the case of zero magnetic field, we are interested in the spectrum to lowest order in $n$, which amounts to expanding the spectrum (\ref{eq:LLn})
in the weak-field limit $\hbar\Omega n/\Delta$, $\hbar\delta\omega/\Delta$, $\hbar\omp\sqrt{n}/\Delta\ll 1$, and one obtains
\beq\label{eq:LLappr}
\epsilon_{\lambda,n}\simeq \lambda\left[\Delta + \hbar\omega_{\lambda}\left(n+\gamma_{\lambda}\right)\right],
\eeq
One first notices that the cyclotron frequencies are indeed those associated with the full band
masses (\ref{eq:bandM}), $\ome=eB/\me$ and $\omh=eB/\mh$, and not given by the bare masses
$\me^0$ and $\mh^0$. The information about the Diracness is encoded in the
phase offset
\beq\label{eq:offset}
\gamma_{\lambda}=\frac{1}{2}(1+\lambda \delta_{\lambda})
\eeq
in terms of the parameter
\beq\label{eq:diracness}
\delta_{\lambda}=\frac{m_{\lambda}^0}{m_D + m_{\lambda}^0 }=\frac{m_{\lambda}}{m_D},
\eeq
which we propose as a measure for the Diracness. Indeed, $\delta_{\lambda}$ varies from zero for Schr\"odinger fermions in the case of fully decoupled bands
to one for pure (massive) Dirac fermions. Notice that, for the expression (\ref{eq:offset}) to be valid in this form, we have rearranged the counting of the
Landau levels such that $n$ starts from zero both in the conduction and in the valence band (see Fig. \ref{fig1}), 
whereas in the expression (\ref{eq:LLn}) the level $n=0$ only
occurs in the valence band, as mentioned above.
The phase offset (\ref{eq:offset}) is thus zero (for holes) or one (for electrons) in the case of
pure Dirac fermions [Fig. \ref{fig1}(c)], whereas one retrieves the usual value $\gamma_{\lambda}=1/2$ for pure Schr\"odinger particles [Fig. \ref{fig1}(a)].
Therefore, the phase offset is no longer necessarily quantized, as shown for an intermediate value in Fig. \ref{fig1}(b),
similarly to the case of surface states in three-dimensional
topological insulators with broken electron-hole symmetry \cite{wright}. 
However, the present treatment shows that the phase offset can
be unquantized even for a particle-hole-symmetric system, $\epsilon_{\lambda}(q)=-\epsilon_{-\lambda}(q)$ in Eq. (\ref{eq:disp00}).
Notice that this symmetry is ensured by the existence of a \textit{chiral} operator (here $i\sigma^y\mathcal{K}$, 
in terms of the complex conjugation $\mathcal{K}$
and the Pauli matrix $\sigma^y$) that anticommutes with the Hamiltonian (\ref{eq:ham0}) if $1/\mu=0$ and $1/\me^0=1/\mh^0=1/M$. The
particle-hole symmetric case is precisely the one depicted in Fig. \ref{fig1}.

\subsection{Semiclassical treatment}

In this subsection, we discuss the Diracness within a semiclassical treatment,
where the reciprocal-space area $S(\epsilon_{\lambda})$ enclosed by the trajectory $\mathcal{C}(\epsilon_{\lambda})$ is quantized via the 
Onsager relation \cite{onsager}
\beq\label{eq:semicl}
S(\epsilon_{\lambda})l_B^2=2\pi(n+\gamma_{\lambda}).
\eeq
The phase offset encodes, by virtue of Eq. (\ref{eq:offset}), the Diracness, which can be related to the Berry phase
\beq\label{eq:Gamma}
\Gamma=\int_{\mathcal{C}(\epsilon_{\lambda})}d\bq\cdot\vec{\Amath}_{\lambda}(\bq)
\eeq
via the relation \cite{roth,fuchs},
\beq\label{eq:diracnessSC}
\delta_{\lambda}=-\frac{\lambda}{\pi}\frac{d(\epsilon\Gamma)}{d|\epsilon_{\lambda}|}=
-\frac{\lambda}{\pi}\frac{d(\epsilon\Gamma)}{d\epsilon}\frac{d\epsilon}{d|\epsilon_{\lambda}|}.
\eeq
In Eq. (\ref{eq:Gamma}) the Berry phase is expressed in terms of the Berry connection
\beq\label{eq:connect}
\vec{\Amath}_{\lambda}(\bq)=i\psi_{\lambda}(\bq)^{\dagger}\nabla_{\bq}\psi_{\lambda}(\bq)=
-\lambda\frac{\epsilon-\Delta-\frac{\hbar^2q^2}{2M}}{2\epsilon}\frac{\be_{\theta}}{q}
\eeq
integrated over the reciprocal-space orbit $\mathcal{C}(\epsilon_{\lambda})$
of energy $\epsilon_{\lambda}$. In the above expressions, $\psi_{\lambda}(\bq)$ are the (spinorial) eigenstates
of Hamiltonian (\ref{eq:ham0}) and $\be_{\theta}$ is the unit vector perpendicular to the wave vector $\bq$. The energy $\epsilon$
that intervenes in the calculation of the phase
offset in Eq. (\ref{eq:semicl}) is that (\ref{eq:disp00}) for $1/\mu=0$, $\epsilon=|\epsilon_{\lambda,1/\mu=0}|$.
The difference between
$\epsilon=\Delta+\hbar^2q^2/2m_{\rm sym}$ and
$|\epsilon_{\lambda}|=\Delta+\hbar^2q^2/2m_{\lambda}$ stems from the electron-hole asymmetry (for $1/\mu\neq 0$), which does not affect the 
eigenstates and the Berry curvature 
but that shifts the energy
of the contour at which the Berry phase $\Gamma$ is evaluated \cite{wright}. 
Whereas the full band mass $m_{\lambda}$ is given in Eq. (\ref{eq:bandM}), the mass $m_{\rm sym}$
entering the energy $\epsilon$ is the symmetric part of the band mass $m_{\rm sym}=(1/M+1/m_D)^{-1}$.

Let us first use the semiclassical approach to analyze the Diracness within the parabolic approximation (\ref{eq:disp0}) for the bands.
From the above expressions one obtains for the Berry connection (\ref{eq:connect})
\beq
\vec{\Amath}_{\lambda}(\bq)=-\lambda\frac{\hbar^2}{4m_D\epsilon}q\be_{\theta}
\eeq
and thus for the Berry curvature $\vec{\Bmath}_{\lambda}(\bq)=\nabla_{\bq}\times\vec{\Amath}_{\lambda}(\bq)=\Bmath_{\lambda}(\bq){\bf e}_z$,
which is a vector oriented in the $z$-direction with
the norm
\beq
\Bmath_{\lambda}(\bq)=\frac{1}{q}\frac{d}{dq}\left[q\Amath_{\lambda}(\bq)\right]\simeq -\lambda\frac{\hbar^2}{2m_D\epsilon},
\eeq
where we have used the fact that the Berry connection is isotropic in reciprocal space, and we have neglected terms of order
$\mathcal{O}[(\epsilon-\Delta)/\Delta]\sim \hbar^2q^2/2m_{\rm sym}\Delta\ll 1$.
Eventually, one obtains from Eq. (\ref{eq:Gamma}) via integration on a path of constant $q(\epsilon)$,
\beq
\Gamma=-\frac{\lambda\pi\hbar^2}{m_D\epsilon}q^2(\epsilon)=\frac{2\pi m_{\rm sym}\Bmath_{\lambda}(\bq=0)}{\hbar^2}\frac{\Delta}{\epsilon}(\epsilon-\Delta),
\eeq
where we have substituted $q^2(\epsilon)=2m_{\rm sym}(\epsilon-\Delta)/\hbar^2$. This yields
\beq
\frac{1}{\pi}\frac{d(\epsilon\Gamma)}{d\epsilon}=\frac{2m_{\rm sym}\Delta\Bmath_{\lambda}(\bq=0)}{\hbar^2},
\eeq
which is the first ingredient in the calculation of the Diracness (\ref{eq:diracnessSC}).
In the last two expressions, we have used the expression
\beq\label{eq:Berry0}
\Bmath_{\lambda}(\bq=0)=-\lambda \frac{\hbar^2}{2m_D\Delta}=-\lambda \frac{\hbar^2v_D^2}{2\Delta^2}.
\eeq
for the Berry curvature directly at the wave-vector position ($\bq=0$) of the gap.
Furthermore, one has $\epsilon=|\epsilon_{\lambda}|-\lambda\hbar^2q^2/2\mu$,
such that $d\epsilon/d|\epsilon_{\lambda}|=1-\lambda m_{\lambda}/\mu$, and the final expression for the Diracness therefore reads
\beq\label{eq:diracnessSCfinal}
\delta_{\lambda}=-\lambda\frac{2m_{\lambda}\Delta\Bmath_{\lambda}(\bq=0)}{\hbar^2}=\frac{2m_{\lambda}\Delta|\Bmath_{\lambda}(\bq=0)|}{\hbar^2}=\frac{m_{\lambda}}{m_D} ,
\eeq
in terms of the full band mass $m_{\lambda}$ of Eq. (\ref{eq:bandM}), which coincides with the Diracness (\ref{eq:diracness}) obtained from the Landau-level
spectrum (\ref{eq:LLn}). The semiclassical treatment thus shows that the Diracness is directly related to the Berry curvature at the gap (at $\bq=0$).
The latter can
be obtained from a numerical determination of the eigenstates in the vicinity of the direct gap ($\bq=0$) within \textit{ab initio} calculations that
also yield the band mass $m_{\lambda}$.

\begin{figure}
\centering
\includegraphics[width=8.5cm,angle=0]{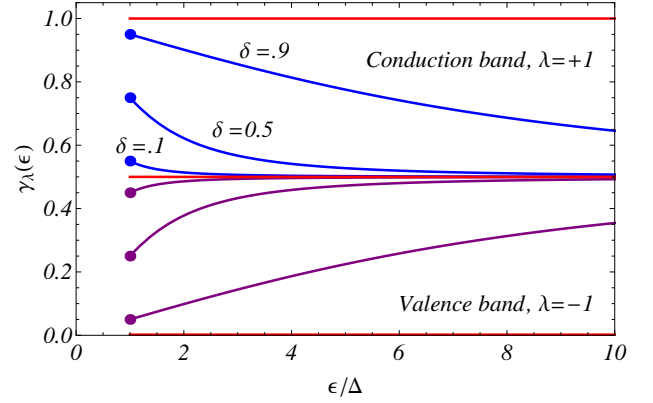}
\caption{\footnotesize{(Color online) Variation of the phase mismatch $\gamma_\lambda(\epsilon)$ as a function of the dimensionless energy $\epsilon/\Delta$, 
in the special case of electron-hole symmetry ($m_e^0=m_h^0=M$), for different values of the Diracness $\delta=M/(M+m_D)$ introduced in Eq. 
(\ref{eq:diracness}). 
The blue lines represent the variation of $\gamma_{\lambda=+1}$ in the conduction band and the purple lines represent the variation of 
$\gamma_{\lambda=-1}$ in the valence band. 
The dots correspond to the low-energy parabolic approximation [Eq. (\ref{eq:LLappr})]. 
When the energy increases the Diracness is reduced and the phase mismatch converges 
towards the Schr\"odinger value $1/2$.
}}
\label{fig2}
\end{figure}

Notice finally that the expression (\ref{eq:diracnessSCfinal}) is valid only in the parabolic-band approximation (\ref{eq:disp0}) and fails at high energies. If
we investigate the model (\ref{eq:ham0}), one may still use the semiclassical approach (\ref{eq:semicl}) together with the expression (\ref{eq:Gamma}) for
the Berry curvature, and the Diracness can be expressed as
\beq
\delta_{\lambda}=-\frac{\lambda}{\pi}\frac{d(\epsilon\Gamma)}{d\epsilon}=1-\frac{\hbar^2}{2M}\frac{dq^2(\epsilon)}{d\epsilon},
\eeq
where we consider for simplicity an electron-hole symmetric band structure with $1/\mu=0$. From the full dispersion relation (\ref{eq:disp00}), one then finds
an energy-dependent Diracness
\beq
\delta_{\lambda}(\epsilon)=1-\frac{\epsilon}{\sqrt{\epsilon^2+M^2v_D^4+2\Delta Mv_D^2}}.
\eeq
Whereas one retrieves the result (\ref{eq:diracness}) in the low-energy limit with $\epsilon\rightarrow \Delta$, one notices that at high energies
the Diracness vanishes $\delta_{\lambda}(\epsilon\gg Mv^2,\Delta)\rightarrow 0$. Indeed, one sees from the dispersion relation (\ref{eq:disp00}) that
the electrons are governed at high energies by the Schr\"odinger terms $\propto \hbar^2q^2/2M$.
The resulting energy-dependent phase offset is plotted in Fig. \ref{fig2}, for different values of the Diracness $\delta=\delta_{\lambda}(\epsilon=\Delta)$
defined within the parabolic approximation (\ref{eq:LLappr}).

\section{General model in the Luttinger-Kohn representation}

Let us now consider the general model of a 2D direct-gap semiconductor that is described within the Luttinger-Kohn representation \cite{LKrep}
to second order in the wave vectors
\beq\label{eq:HamLK}
H=\left(
\begin{array}{cc}
\Delta + \frac{\hbar^2}{2m_{ij}^{\rm e}}q_i q_j & \hbar(\bv_1\cdot \bq -i \bv_2\cdot \bq)\\
\hbar(\bv_1\cdot \bq + i \bv_2 \cdot \bq) & -\Delta - \frac{\hbar^2}{2m_{ij}^{\rm h}}q_i q_j
\end{array}
\right),
\eeq
where a summation over repeated indices $i,j$ is implicit. There are no terms linear in $\bq$ in the diagonal components because we require, for a
direct-gap semiconductor, both a minimum in the conduction band and a maximum of the valence band at $\bq=0$.
One notices that the band structure is now anisotropic, and
the principle axes for the ellipses in the conduction band are generally not identical to those in the valence band.

\subsection{Berry curvature}

The Berry curvature $\Bmath_{\lambda}(\bq)$ is a property encoded in the Hamiltonian or else the
eigenstates $\psi_{\lambda}(\bq)$ of the Hamiltonian (\ref{eq:HamLK}),
\beq
\Bmath_{\lambda}(\bq) = -\frac{\lambda}{2|\vec{d}_{\bq}|^3}\vec{d}_{\bq}\cdot\left(\partial_{q_x}\vec{d}_{\bq}\times \partial_{q_y}\vec{d}_{\bq}\right)
\label{eq:Berry}
\eeq
where we have made use of a gauge-invariant expression \cite{refBerryCurv} in terms of the vector $\vec{d}_{\bq}$ the components of which
read $d_{\bq}^x=\hbar\bv_1\cdot\bq$, $d_{\bq}^y=\hbar\bv\cdot\bq$, and $d_{\bq}^z=\Delta +(\hbar^2/2M_{ij})q_iq_j$. Here, the mass tensors
$1/m_{ij}^{\rm e} = 1/M_{ij} + 1/\mu_{ij}$ $1/m_{ij}^{\rm h}=1/M_{ij} - 1/\mu_{ij}$ are written in terms of an average mass $M_{ij}$ and the
mass anisotropy $\mu_{ij}$, which does not enter in the calculation of the Berry curvature. This calculation is straight-forward if one chooses a frame
of reference which coincides with the principle axes of the tensor $1/M_{ij}$ such that the latter is diagonal. Within the semiclassical treatment,
the relevant quantity is the Berry curvature directly at the band gap, and one obtains at $\bq=0$ a generalized version of Eq. (\ref{eq:Berry0})
\beq\label{eq:BerryLK}
\mathcal{B}_{\lambda}(\bq=0)=-\lambda\frac{\hbar^2\bv_1\wedge\bv_2}{2\Delta^2}=-\lambda\frac{\hbar^2}{2\Delta m_D}.
\eeq
Here, the Dirac mass reads
\beq\label{eq:DiracMassLK}
m_D=\frac{\Delta}{\bv_1\wedge\bv_2},
\eeq
in terms of the 2D vector product $\bv_1\wedge\bv_2\equiv (\bv_1\times\bv_2)_z=v_1^xv_2^y-v_1^yv_2^x$,
which is the $z$-component of the 3D vector product.
Notice, however, that the expression (\ref{eq:BerryLK}) for the Berry curvature
is valid strictly speaking only in a two-band model, where the wave functions have two components. In the presence of additional bands, it cannot be excluded
that they contribute to the Berry curvature of the bands $\lambda={\rm e,\,h}$.

\subsection{Cyclotron frequency and Diracness in the conduction band}

The Landau-level spectrum associated with Hamiltonian (\ref{eq:HamLK}) can be calculated approximately in the large-gap limit.
We concentrate on the spectrum in the conduction band, that is for energies $\epsilon\sim \Delta$ -- the arguments are easily adapted for the valence band. In this
case, it is convenient to choose a frame of reference for the wave vectors that coincides with the principle axes of the electron mass tensor $1/m_{ij}^{\rm e}$
(for a calculation of the spectrum in the valence band, one would have chosen a diagonal tensor $1/m_{ij}^{\rm h}$),
and the Peierls substitution yields
\beq
\frac{\hbar^2}{2m_{ij}^{\rm e}}q_iq_j=\frac{\hbar^2 q_x^2}{2m_x^{\rm e}}+ \frac{\hbar^2 q_y^2}{2m_y^{\rm e}}\quad \rightarrow\quad \hbar\ome^0(a^{\dagger}a +1/2),
\eeq
where $\ome^0=eB/\me^0$, with the average mass $\me^0=\sqrt{m_x^{\rm e}m_y^{\rm e}}$. The ladder operator reads $a=l_B(\bar{q}_x - i\bar{q}_y)/\sqrt{2}$, in
terms of the rescaled wave-vector components $\bar{q}_x=\sqrt{\me^0/m_x^{\rm e}}q_x$ and $\bar{q}_y=\sqrt{\me^0/m_y^{\rm e}}q_y$. 
Hamiltonian (\ref{eq:HamLK}) can thus be rewritten as
\beq
H=\left(
\begin{array}{cc}
\Delta + \hbar\ome^0(a^{\dagger}a + 1/2) & \hbar(\omega'a+\omega''a^{\dagger})\\
\hbar({\omega'}^{*}a^{\dagger} + {\omega''}^*a) & -\Delta +\mathcal{O}
\end{array}
\right),
\eeq
where $\mathcal{O}=- \hbar\omh(a^{\dagger}a +1/2) +\hbar\omh'a^{\dagger2} + \hbar\omh'' a^2\ll \Delta$ represents terms that are neglected in the large-gap limit when
considering Landau levels in the conduction band. The other cyclotron-type frequencies are
\beqn\nn
\omega'  &=& \frac{1}{\sqrt{2}l_B}\left(\bar{v}_1 - i\bar{v}_2\right)\\
\omega'' &=& \frac{1}{\sqrt{2}l_B}\left(\bar{v}_1^* - i\bar{v}_2^*\right),
\eeqn
in complex notation with the rescaled velocities $\bar{v}_j=\sqrt{m_x^{\rm e}/\me^0}\, v_j^x + i\sqrt{m_y^{\rm e}/\me^0}\, v_j^y$.
The solution of the eigenvalue equation $H\psi_n = \epsilon_n\psi_n$ in the conduction band with the spinorial form $\psi_n=(u_n,v_n)$ yields the equation
\beqn
&&\left[(\Delta -\epsilon_n) + \hbar\ome^0(a^{\dagger}a+1/2) \right. \\
\nn
&&+\left.
\hbar^2\frac{|\omega'|^2aa^{\dagger}+|\omega''|^2a^{\dagger}a + \omega'{\omega''}^*a^2 +{\omega'}^*\omega''a^{\dagger 2}}{2\Delta}\right]u_n\simeq 0
\eeqn
for the large component $u_n$, which can be diagonalized with the help of the canonical transformation $a=\sinh\beta \exp(-i\phi) b^{\dagger} + \cosh\beta b$,
such as to get rid of the terms proportional to $bb$ and $b^{\dagger}b^{\dagger}$. 
A lengthy but straight-forward calculation then yields the Landau-level spectrum
\beq
\epsilon_n \simeq \Delta + \hbar \ome( n +\gamma_{\rm e}),
\eeq
where
$\ome = eB/\me^C$
is the cyclotron frequency in terms of the cyclotron mass

\beq\label{eq:cyclM}
\me^C=\left[\frac{1}{(\me^0)^2}+\frac{1}{\me^0}\frac{|\bar{v}_1|^2+|\bar{v}_2|^2}{\Delta}+\frac{1}{m_D^2}\right]^{-1/2}
\eeq
for the electrons in the conduction band. The cyclotron mass (\ref{eq:cyclM}) coincides with the expression (\ref{eq:bandM}) for the electronic band
mass in the simplified model for an isotropic bare mass $\me^0$ and an isotropic Dirac velocity $\bv_1=(v_D,0)$ and $\bv_2=(0,v_D)$.
The shift can again be related by Eq. (\ref{eq:offset}), $\gamma_{\rm e}=(1+\delta_{\rm e})/2$, to the Diracness
\beq\label{eq:diracnessLK}
\delta_{\rm e}=\frac{\hbar \bv_1\wedge\bv_2}{\Delta \ome l_B^2}=\frac{\me^C}{m_D}=-\frac{2\Delta \mathcal{B}_{\rm e}(\bq=0)\me^C}{\hbar^2},
\eeq
written in terms of the cyclotron mass (\ref{eq:cyclM}) and the Dirac mass (\ref{eq:DiracMassLK}).
One thus retrieves also in the Luttinger-Kohn model (\ref{eq:HamLK})
the same parameter for the Diracness as in the simplified model (\ref{eq:ham0}). We have expressed, in the last step, the Diracness in terms
of the Berry curvature $\mathcal{B}_{\rm e}(\bq=0)$ at the direct gap via Eq. (\ref{eq:BerryLK}),
such that one finds the same expression (\ref{eq:diracnessSCfinal}) as in the
semiclassical approach in the parabolic-band approximation. 

Notice that one obtains a Diracness that can be positve or negative depending on the sign of 
$\bv_1\wedge\bv_2$. As in the simplified model discussed above, the sign indicates whether the Landau level $n=0$ resides in
the conduction or the valence band. Furthermore, it is apparent from Eq. (\ref{eq:diracnessLK}) that the Diracness as well as the 
Berry curvature (\ref{eq:BerryLK}) at the gap are annihilated if the two velocities $\bv_1$ and $\bv_2$ are collinear. Indeed, in the absence of the mass terms 
$1/m_{ij}^{\rm e}$, the dispersion would not depend on the wave vector in the direction perpendicular to $\bv_1$ and $\bv_2$ such that there would be no closed orbits
and consequently no Landau levels. The Landau-level spectrum in this case therefore arises solely from the parabolic terms, and the electrons are of Schr\"odinger
type with $\delta_{\rm e}=0$.

\section{Conclusions}

In conclusion, we have shown a general path to quantify the Dirac character of electrons in 2D direct-gap semiconductors. The Diracness, which may be viewed
as the contribution of the Dirac mass 
$m_D=\Delta/\bv_1\wedge\bv_2=\Delta/(v_1^xv_2^y-v_1^yv_2^x)$
to the full band mass, is not revealed directly by the band structure of the electronic
system. This subtle parameter is indeed encoded in the form of the two-band model, which finds its most general expression in the Luttinger-Kohn representation.
It is unveiled in the phase offset of Shubnikov-de-Haas oscillations. Also in \textit{ab initio} calculations, the Diracness may be extracted from the calculation of
the Berry curvature directly at the gap. Due to its physical content, the proposed identification of the Diracness is an essential ingredient for the understanding
of a vast second generation of 2D crystals, beyond graphene, with semiconducting electrons.

The authors thank Jean-No\"el Fuchs for fruitful discussions and a careful reading of the manuscript.


\begin{thebibliography}{0}
\bibitem{note1}
Notice that
we use the term ``Dirac velocity'' to denote a parameter in this description. It coincides with the Fermi velocity only in the case of a vanishing gap
$\Delta$ and in the absence of parabolic terms, $1/\me^0=1/\mh^0=0$.
\end{thebibliography}
\end{document}